\documentclass[letterpaper, 10 pt, conference]{ieeeconf} 
\IEEEoverridecommandlockouts  
\usepackage{amsmath}
\usepackage{amssymb}
\usepackage{mathtools}
\usepackage[final]{microtype}
\usepackage{graphicx}
\usepackage{algorithm}
\usepackage{algorithmic}
\usepackage{booktabs}
\usepackage{bm}
\usepackage{multirow}

\usepackage{subcaption}
\usepackage{hyperref}
\usepackage{url}
\usepackage{xcolor}
\hypersetup{
    colorlinks=true,
    linkcolor=blue,
    citecolor=blue,
    urlcolor=blue
}

\newtheorem{theorem}{Theorem}
\newtheorem{lemma}[theorem]{Lemma}
\newtheorem{proposition}[theorem]{Proposition}

\newtheorem{definition}[theorem]{Definition}
\newtheorem{remark}{Remark}
\newtheorem{assumption}[theorem]{Assumption}

\title{\LARGE \bf
Smart Predict-Then-Control: \\Control-Aware Surrogate Refinement for System Identification }

\author{
Jiachen Li$^{1}$, Shihao Li$^{1}$, and Dongmei Chen$^{1}$%
\thanks{$^{1}$All authors are with the Department of Mechanical Engineering, The University of Texas at Austin, Austin, TX 78712, USA.
{\tt\small \{jiachenli, shihaoli01301, dmchen\}@utexas.edu}}%
}
\begin{document}

\maketitle
\thispagestyle{empty}
\pagestyle{empty}

\begin{abstract}
This paper introduces Smart Predict-Then-Control (SPC), a control-aware refinement procedure for model-based control. SPC refines a prediction-oriented model by optimizing a surrogate objective that evaluates candidate models through the control actions they induce. For a fixed-surrogate variant under unconstrained control, we establish the smoothness of the surrogate, projected-gradient convergence at the $O(1/K)$ rate, and a bias decomposition that yields a conditional transfer diagnostic. On a wind-disturbed quadrotor trajectory-tracking task, Updated-SPC reduces tracking RMSE by 70\% and closed-loop cost by 42\% relative to the nominal baseline.
\end{abstract}

\section{INTRODUCTION}

When system dynamics are known precisely, constructing a near-optimal controller is largely a solved problem. In practice, however, model uncertainty, measurement noise, and unmodeled dynamics conspire to make such knowledge elusive. System identification bridges this gap by fitting a model from data—typically by minimizing prediction error. But a model that excels at prediction need not excel at control: small errors along parameter directions to which the controller is sensitive can disproportionately degrade the induced control law \cite{gevers2005, bristow2006}. The root cause is a misalignment between the identification objective and the downstream control task. Prediction-error minimization distributes its fitting effort uniformly across all output directions, whereas the controller may depend critically on a small subset of model parameters. Consequently, the best predictor and the best model for control are, in general, different.

A natural response is to inject control-relevant information into the identification stage \cite{gevers2005, hjalmarsson2005}. Existing approaches do so in two ways: one modifies the prediction loss with control-oriented penalties \cite{formentin2021}, and the other abandons modularity entirely, training the model end-to-end through a differentiable controller \cite{amos2018, drgona2022}. Both have limitations. The former typically yields a reweighted prediction loss that may not capture the full sensitivity structure of the control map; the latter sacrifices interpretability and safety guaranties \cite{brunke2022}. What is missing is an approach that retains modularity while optimizing an objective that reflects how a candidate model will perform once a controller is designed around it.

\textbf{Related Work.} The tension between prediction accuracy and closed-loop performance has been recognized since the early identification-for-control literature \cite{gevers2005, hjalmarsson2005}. Control-oriented regularization \cite{formentin2021} augments the identification loss with control-relevant priors but remains limited to linear systems. Goal-driven dynamics learning \cite{bansal2017} employs Bayesian optimization over a control cost yet provides no surrogate with tractable analytical properties. Differentiable MPC \cite{amos2018, drgona2022} and actor-critic MPC \cite{romero2024} permit end-to-end training at the expense of interpretability and safety \cite{brunke2022}. In the optimization literature, the Smart Predict-then-Optimize (SPO) framework \cite{elmachtoub2022} aligns predictions with decision outcomes for combinatorial problems; recent extensions to control \cite{cui2025, favaro2025} adopt task-specific losses with application-dependent theory. Lin et al. \cite{lin2025} formalize the prediction–control misalignment for LQR but stop short of constructing an improved identification procedure. Yu et al. \cite{yu2020} study the role of predictions in online LQR.

\textbf{Positioning and Contributions.} This paper proposes Smart Predict-Then-Control (SPC), which addresses the limitations above by constructing and optimizing a \emph{surrogate objective} that evaluates candidate models through the control actions they induce. Unlike control-oriented regularization, the SPC surrogate is not a reweighted prediction loss: it couples model evaluation with the control optimization map, so that the gradient signal reflects how parameter changes propagate through the optimizer into control performance (see Remark~\ref{rem:not_reweighting}). Unlike end-to-end approaches, SPC preserves the modular predict-then-control pipeline.

Theoretically, for the fixed-surrogate variant under unconstrained control, we establish surrogate smoothness, projected-gradient convergence at the $O(1/K)$ rate, and a bias-decomposition-based conditional transfer result linking surrogate improvement to deployment performance. Empirically, both SPC variants substantially reduce tracking error on a wind-disturbed quadrotor task relative to nominal, control-weighted regularization, and differentiable-controller baselines. The updated-surrogate variant and the constrained MPC used in experiments are practical extensions beyond the current theory.

\section{PROBLEM FORMULATION}

Consider a discrete-time dynamical system parameterized by $\theta \in \Theta \subseteq \mathbb{R}^p$ over a finite horizon $T$:
\begin{equation}
\label{eq:dynamics}
x_{t+1}=f(x_t,u_t;\theta)+w_t,
\qquad t=0,1,\dots,T-1,
\end{equation}
where $x_t\in\mathbb{R}^n$ is the state, $u_t\in\mathbb{R}^m$ the control input, $f:\mathbb{R}^n\times\mathbb{R}^m\times\Theta\to\mathbb{R}^n$ a differentiable dynamics map, $w_t\in\mathbb{R}^n$ an unknown disturbance, and $x_0$ a given initial condition.

The theoretical development assumes \emph{unconstrained} control sequences $U=(u_0,\dots,u_{T-1})\in\mathbb{R}^{mT}$.

For any initial state $x_0$, control sequence $U=(u_0,\dots,u_{T-1})$, parameter $\theta$, and disturbance sequence $W=(w_0,\dots,w_{T-1})$, let $\{x_t(x_0,U,\theta,W)\}_{t=0}^T$ denote the trajectory generated by \eqref{eq:dynamics}. Writing $x_t\equiv x_t(x_0,U,\theta,W)$ for brevity, the finite-horizon cost is
\begin{equation}
\label{eq:cost_explicit}
J(x_0,U;\theta,W)=
\sum_{t=0}^{T-1}\bigl(x_t^\top Qx_t+u_t^\top Ru_t\bigr)+x_T^\top Px_T,
\end{equation}
with $Q,P\succeq 0$ and $R\succ 0$.

\begin{definition}[Deployment metric]
\label{def:deploy}
For a parameter estimate $\theta\in\Theta$, define
\begin{equation}
\label{eq:deploy}
\begin{aligned}
V(\theta)
={}&\frac{1}{N_{\mathrm{dep}}}\sum_{i=1}^{N_{\mathrm{dep}}}
J\bigl(
 x_{0,i}^{\mathrm{dep}},
 U^*(x_{0,i}^{\mathrm{dep}};\theta);\\
&\hspace{4.9em}\theta_{\mathrm{true}},W_i^{\mathrm{true}}
\bigr).
\end{aligned}
\end{equation}
where $U^*(x_0;\theta)$ denotes the scenario-optimal control induced by the identified model $\theta$ for the initial state $x_0$, and $\{(x_{0,i}^{\mathrm{dep}},W_i^{\mathrm{true}})\}_{i=1}^{N_{\mathrm{dep}}}$ are deployment scenarios.
\end{definition}

$V(\theta)$ serves as an analytical reference metric; the SPC algorithm requires access to neither $\theta_{\mathrm{true}}$ nor deployment scenarios. The practical evaluation in Section~\ref{sec:experiments} employs receding-horizon MPC on a simulated quadrotor trajectory-tracking task, which differs from the open-loop formulation in \eqref{eq:deploy}.

\section{METHOD}

\begin{figure}[t]
\centering
\includegraphics[width=\columnwidth]{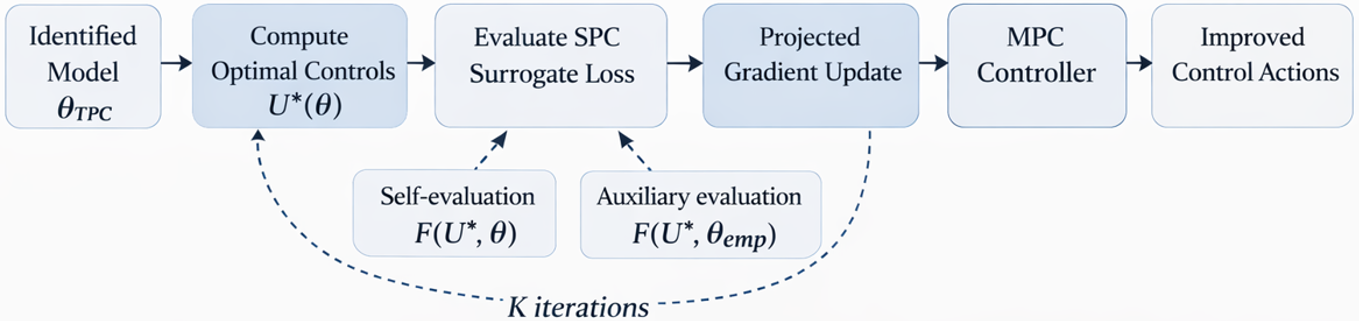}
\caption{Overview of the SPC refinement pipeline.}
\label{fig:spc_overview}
\end{figure}

Figure~\ref{fig:spc_overview} illustrates the SPC pipeline. 

The dataset $\mathcal D=\{(X_i,U_i)\}_{i=1}^N$ contains $N$ state-control trajectories, split into training ($\mathcal D_{\mathrm{tr}}$, 80\%) and test ($\mathcal D_{\mathrm{te}}$, 20\%) sets with sizes $N_{\mathrm{tr}}$ and $N_{\mathrm{te}}$, respectively. Identification and surrogate optimization use only $\mathcal D_{\mathrm{tr}}$; all closed-loop metrics are reported on $\mathcal D_{\mathrm{te}}$.

\subsection{Initialization}

The procedure begins from the prediction-oriented estimator
\begin{equation}
\label{eq:init}
\theta^{(0)}=\theta_{\mathrm{TPC}}
=
\arg\min_{\theta\in\Theta}
\sum_{i\in\mathcal D_{\mathrm{tr}}}\sum_{t=0}^{T-1}
\|x_{t+1,i}-f(x_{t,i},u_{t,i};\theta)\|_2^2.
\end{equation}

From $\theta^{(0)}$, we construct residual-based disturbance scenarios
\begin{equation}
\label{eq:dist_est}
w_{t,i}
=
x_{t+1,i}-f(x_{t,i},u_{t,i};\theta^{(0)}),
\end{equation}
collected into $W_i=(w_{0,i},\dots,w_{T-1,i})$. These residuals are computed once and held fixed throughout; they serve as surrogates for the unmodeled disturbance rather than independent disturbance estimates.

\subsection{Auxiliary Evaluation Parameter}

The theoretical development requires only an auxiliary parameter $\theta_{\mathrm{emp}}\in\Theta$ that remains fixed during the SPC iterations. No statistical consistency of $\theta_{\mathrm{emp}}$ is assumed or needed for the convergence results that follow.

When a data-driven instantiation is desired, it must be constructed from counterfactual rollouts that are internally consistent with the optimized controls. Concretely, after obtaining $\theta^{(0)}=\theta_{\mathrm{TPC}}$ and the residual scenarios $\{W_i\}$ from \eqref{eq:dist_est}, define $\hat X_i^{(0)}=\{\hat x_{t,i}^{(0)}\}_{t=0}^T$ by
\begin{equation}
\label{eq:counterfactual_rollout}
\begin{aligned}
\hat x_{0,i}^{(0)} &= x_{0,i},\\
\hat x_{t+1,i}^{(0)}
&=f\bigl(\hat x_{t,i}^{(0)},u_{t,i}^*(\theta^{(0)});\theta^{(0)}\bigr)
+w_{t,i}.
\end{aligned}
\end{equation}
A key distinction: the counterfactual rollout applies the optimized controls $u_{t,i}^*(\theta^{(0)})$ rather than the original recorded controls $u_{t,i}$ appearing in the residual computation \eqref{eq:dist_est}. The residuals approximate the unmodeled disturbance under the data-collection policy, while the rollout evaluates the model along the optimized trajectory.

Here $u_{t,i}^*(\theta^{(0)})$ denotes the $t$-th component of $U_i^*(\theta^{(0)})$. One coherent choice is then
\begin{equation}
\label{eq:emp_init_revised}
\begin{aligned}
\theta_{\mathrm{emp}}
=\arg\min_{\theta\in\Theta}
\sum_{i\in\mathcal D_{\mathrm{tr}}}\sum_{t=0}^{T-1}
\bigl\|&\hat x_{t+1,i}^{(0)} \\
&-f\bigl(\hat x_{t,i}^{(0)},u_{t,i}^*(\theta^{(0)});\theta\bigr)-w_{t,i}
\bigr\|_2^2.
\end{aligned}
\end{equation}
In the theory section, however, $\theta_{\mathrm{emp}}$ is treated simply as a fixed element of $\Theta$; none of the properties of this particular construction are invoked.

\subsection{Scenario-Based Control Computation}

For each trajectory $i$ and parameter $\theta$, define the scenario objective
\begin{equation}
\label{eq:Fi_method}
F_i(U,\theta)
\triangleq
\frac{1}{N_{\mathrm{tr}}}\sum_{j\in\mathcal D_{\mathrm{tr}}}
J(x_{0,i},U;\theta,W_j),
\end{equation}
where each term evaluates the cost under the same initial state $x_{0,i}$ and control sequence $U$, but with model parameter $\theta$ and disturbance sequence $W_j$. The scenario-optimal control is
\begin{equation}
\label{eq:mpc}
U_i^*(\theta)
=
\arg\min_{U\in\mathbb{R}^{mT}} F_i(U,\theta).
\end{equation}
This is the scenario-level specialization of $U^*(x_0;\theta)$ from Definition~\ref{def:deploy}: $U_i^*(\theta)\equiv U^*(x_{0,i};\theta)$, with the initial state $x_{0,i}$ implicit in the index~$i$.

\subsection{SPC Surrogate Loss}

The fixed-surrogate objective takes the form
\begin{equation}
\label{eq:tildeL_method}
\widetilde{\mathcal L}(\theta;\theta_{\mathrm{emp}})
\triangleq
\frac{1}{N_{\mathrm{tr}}}\sum_{i\in\mathcal D_{\mathrm{tr}}}
\Bigl[
2F_i(U_i^*(\theta),\theta)
-
F_i(U_i^*(\theta),\theta_{\mathrm{emp}})
\Bigr].
\end{equation}
The term $F_i(U_i^*(\theta),\theta_{\mathrm{emp}})$ evaluates the control induced by $\theta$ under the auxiliary model, while $F_i(U_i^*(\theta),\theta)$ penalizes self-evaluation bias. This construction draws on SPO-style surrogate design \cite{elmachtoub2022}, adapted here to the control setting. It is a heuristic control-aware objective rather than a calibrated bound on $V(\theta)$.

\begin{remark}[Not reweighting]
\label{rem:not_reweighting}
Control-weighted regularization (CW-Reg) appends a control-cost penalty to the prediction loss, producing an objective of the form $\mathcal L_{\mathrm{pred}}(\theta)+\lambda\cdot\text{(control cost)}$. The SPC surrogate \eqref{eq:tildeL_method} is structurally different: it evaluates the \emph{optimized} control $U_i^*(\theta)$, which itself depends on $\theta$ through the argmin in \eqref{eq:mpc}. 
\end{remark}

\subsection{Fixed-Surrogate SPC Algorithm}

\begin{algorithm}
\caption{Fixed-Surrogate SPC}
\label{alg:fixed_spc}
\begin{algorithmic}[1]
\STATE \textbf{Input:} $\mathcal D_{\mathrm{tr}}$, step size $\eta$, iterations $K$
\STATE Compute $\theta^{(0)}=\theta_{\mathrm{TPC}}$ via \eqref{eq:init}
\STATE Compute $\{W_i\}$ via \eqref{eq:dist_est} and choose a fixed
$\theta_{\mathrm{emp}}\in\Theta$ (e.g., via \eqref{eq:emp_init_revised})
\FOR{$k=0,\dots,K-1$}
    \STATE Compute $U_i^*(\theta^{(k)})$ via \eqref{eq:mpc} for each $i$
    \STATE $\theta^{(k+1)} \leftarrow \Pi_\Theta\!\bigl[\theta^{(k)}
    - \eta\nabla_\theta\widetilde{\mathcal L}(\theta^{(k)};\theta_{\mathrm{emp}})\bigr]$
\ENDFOR
\STATE \textbf{Output:} $\theta^{(K)}$
\end{algorithmic}
\end{algorithm}

\begin{remark}[Updated-surrogate variant]
\label{rem:practical_spc}
In practice, periodically recomputing $\theta_{\mathrm{emp}}$ every $\tau$ iterations can improve surrogate--deployment alignment. This updated-surrogate variant is a heuristic extension; the convergence and transfer results below apply only to the fixed-surrogate case.
\end{remark}

\section{THEORETICAL ANALYSIS}
\label{sec:theory}

We analyze the fixed-surrogate variant, holding $\theta_{\mathrm{emp}}\in\Theta$ constant throughout the optimization. All results in this section pertain to the unconstrained setting. Proofs are deferred to Appendix.

For any $(x_0,U,\theta,W)$, let $\{x_t(x_0,U,\theta,W)\}_{t=0}^T$ denote the trajectory generated by
\begin{equation}
\label{eq:trajectory_theory}
\begin{aligned}
x_{t+1}(x_0,U,\theta,W)
&=f\bigl(x_t(x_0,U,\theta,W),u_t;\theta\bigr)+w_t,\\
x_0(x_0,U,\theta,W)&=x_0.
\end{aligned}
\end{equation}
Abbreviating $x_t\equiv x_t(x_0,U,\theta,W)$, the finite-horizon cost reads
\begin{equation}
\label{eq:cost_theory}
J(x_0,U;\theta,W)=
\sum_{t=0}^{T-1}\bigl(x_t^\top Qx_t+u_t^\top Ru_t\bigr)+x_T^\top Px_T.
\end{equation}

For each training index $i$, define
\begin{equation}
\label{eq:Fi_theory}
F_i(U,\theta)
\triangleq
\frac{1}{N_{\mathrm{tr}}}
\sum_{j\in\mathcal D_{\mathrm{tr}}}
J(x_{0,i},U;\theta,W_j),
\end{equation}
and
\begin{equation}
\label{eq:Ustar_theory}
U_i^*(\theta)
=
\arg\min_{U\in\mathbb{R}^{mT}} F_i(U,\theta).
\end{equation}
The fixed-surrogate objective is
\begin{equation}
\label{eq:tildeL_theory}
\widetilde{\mathcal L}(\theta;\theta_{\mathrm{emp}})
=
\frac{1}{N_{\mathrm{tr}}}
\sum_{i\in\mathcal D_{\mathrm{tr}}}
\Bigl[
2F_i(U_i^*(\theta),\theta)
-
F_i(U_i^*(\theta),\theta_{\mathrm{emp}})
\Bigr].
\end{equation}

\subsection{Regularity of the Surrogate}

\begin{assumption}[Compact parameter set]
\label{as:theta_compact_revised}
$\Theta\subset\mathbb{R}^p$ is nonempty, compact, and convex.
\end{assumption}

\begin{assumption}[Objective regularity]
\label{as:Fi_reg_revised}
There exists an open set $\mathcal O\supset\Theta$ such that for each $i$:
\begin{enumerate}
    \item $F_i:\mathbb{R}^{mT}\times\mathcal O\to\mathbb{R}$ is $C^3$;
    \item $\nabla_{UU}^2 F_i(U,\theta)\succeq \mu_U I$ for all
$(U,\theta)\in\mathbb{R}^{mT}\times\mathcal O$, where $\mu_U>0$ is independent of $i$.
\end{enumerate}
\end{assumption}

Assumption~\ref{as:Fi_reg_revised} is satisfied by unconstrained finite-horizon LQR with $R\succ 0$. It generally fails for constrained MPC, where the optimizer map can lose smoothness at active-set transitions.

\begin{proposition}[Smoothness of optimizer map/ surrogate]
\label{prop:smoothness_revised}
Under Assumptions~\ref{as:theta_compact_revised}--\ref{as:Fi_reg_revised}, the following hold for each $i$:
\begin{enumerate}
    \item For every $\theta\in\Theta$, the problem
$\min_{U\in\mathbb{R}^{mT}} F_i(U,\theta)$ admits a unique minimizer $U_i^*(\theta)$.
    \item The map $U_i^*:\Theta\to\mathbb{R}^{mT}$ is $C^2$, with
    \begin{equation}
    \label{eq:DUstar}
    D U_i^*(\theta)
    =
    -\Bigl[\nabla_{UU}^2 F_i\bigl(U_i^*(\theta),\theta\bigr)\Bigr]^{-1}
    \nabla_{U\theta}^2 F_i\bigl(U_i^*(\theta),\theta\bigr).
    \end{equation}
    \item For fixed $\theta_{\mathrm{emp}}\in\Theta$,
$\widetilde{\mathcal L}(\cdot;\theta_{\mathrm{emp}})$ is $C^2$ on $\Theta$.
    \item There exists $L_\nabla>0$ such that
    \begin{equation}
    \begin{aligned}
    &\|\nabla \widetilde{\mathcal L}(\theta;\theta_{\mathrm{emp}})
    -\nabla \widetilde{\mathcal L}(\theta';\theta_{\mathrm{emp}})\|\\
    &\qquad \le L_\nabla \|\theta-\theta'\|,
    \qquad \forall \, \theta,\theta'\in\Theta.
    \end{aligned}
    \end{equation}
\end{enumerate}
\end{proposition}

\subsection{Convergence of Projected-Gradient SPC}

Define the standard gradient mapping for the constrained problem:
\begin{equation}
\mathcal G_\eta(\theta)
\triangleq
\frac{1}{\eta}
\left(
\theta
-
\Pi_\Theta\bigl[\theta-\eta\nabla
\widetilde{\mathcal L}(\theta;\theta_{\mathrm{emp}})\bigr]
\right).
\end{equation}

\begin{theorem}[Gradient convergence]
\label{thm:stationarity_revised}
Under Assumptions~\ref{as:theta_compact_revised}--\ref{as:Fi_reg_revised}, let $\{\theta^{(k)}\}_{k\ge 0}$ be the iterates of Algorithm~\ref{alg:fixed_spc}. If $\eta\le 1/L_\nabla$, then for every $K\ge 1$,
\begin{equation}
\label{eq:stationarity_rate_revised}
\min_{0\le k\le K-1}\|\mathcal G_\eta(\theta^{(k)})\|^2
\le
\frac{
2\bigl(
\widetilde{\mathcal L}(\theta^{(0)};\theta_{\mathrm{emp}})
-
\widetilde{\mathcal L}^*
\bigr)
}{
\eta K
},
\end{equation}
where
\[
\widetilde{\mathcal L}^* = \min_{\theta\in\Theta} \widetilde{\mathcal L}(\theta;\theta_{\mathrm{emp}}).
\]
Moreover, every accumulation point of $\{\theta^{(k)}\}$ is first-order stationary for $\min_{\theta\in\Theta}\widetilde{\mathcal L}(\theta;\theta_{\mathrm{emp}})$.
\end{theorem}

\subsection{Bias Decomposition and Conditional Transfer}

Recall the offline deployment metric
\begin{equation}
\label{eq:deploy_theory}
V(\theta)
=
\frac{1}{N_{\mathrm{dep}}}
\sum_{i=1}^{N_{\mathrm{dep}}}
J\bigl(x_{0,i}^{\mathrm{dep}},U^*(x_{0,i}^{\mathrm{dep}};\theta);
\theta_{\mathrm{true}},W_i^{\mathrm{true}}\bigr),
\end{equation}
where $U^*(x_0;\theta)$ denotes the control sequence obtained by solving the scenario optimization problem with initial state $x_0$ and model parameter $\theta$, consistent with $U_i^*(\theta)$ in \eqref{eq:mpc}. The SPC algorithm does not require knowledge of $\theta_{\mathrm{true}}$; $V(\theta)$ enters only as an analytical reference.

Define the bias function
\begin{equation}
\label{eq:bias_def_revised}
B(\theta;\theta_{\mathrm{emp}})
\triangleq
V(\theta)-\widetilde{\mathcal L}(\theta;\theta_{\mathrm{emp}}).
\end{equation}

\begin{proposition}[Bias decomposition]
\label{prop:bias_revised}
For any $\theta,\theta'\in\Theta$ and fixed $\theta_{\mathrm{emp}}\in\Theta$,
\begin{equation}
\label{eq:bias_decomp_revised}
\begin{aligned}
V(\theta)-V(\theta')
={}&\Bigl[\widetilde{\mathcal L}(\theta;\theta_{\mathrm{emp}})
-\widetilde{\mathcal L}(\theta';\theta_{\mathrm{emp}})\Bigr]\\
&+\Bigl[B(\theta;\theta_{\mathrm{emp}})-B(\theta';\theta_{\mathrm{emp}})\Bigr].
\end{aligned}
\end{equation}
\end{proposition}

\begin{assumption}[Bias regularity]
\label{as:bias_reg_revised}
There exists $L_B\ge 0$ such that for all $\theta,\theta'\in\Theta$,
\begin{equation}
|B(\theta;\theta_{\mathrm{emp}})-B(\theta';\theta_{\mathrm{emp}})|
\le
L_B\|\theta-\theta'\|.
\end{equation}
\end{assumption}

\begin{theorem}[Conditional transfer]
\label{thm:transfer_revised}
Under Assumption~\ref{as:bias_reg_revised}, if
\begin{equation}
\label{eq:transfer_cond_revised}
\widetilde{\mathcal L}(\theta^{(0)};\theta_{\mathrm{emp}})
-
\widetilde{\mathcal L}(\theta^{(K)};\theta_{\mathrm{emp}})
>
L_B\|\theta^{(K)}-\theta^{(0)}\|,
\end{equation}
then
\begin{equation}
V(\theta^{(K)})<V(\theta^{(0)}).
\end{equation}
\end{theorem}

\section{EXPERIMENTS}
\label{sec:experiments}

\subsection{Setup}

We evaluate SPC on a wind-disturbed quadrotor trajectory-tracking task. The reference trajectory spans $24$\,s at a sampling interval of $\Delta t=0.02$\,s and comprises three phases: a circular arc of radius $2.5$\,m, an S-shaped descent with a lateral sway of $1.0$\,m down to $z=-2.5$\,m, and a figure-8 with amplitudes of $2.5$\,m in $x$ and $1.25$\,m in $y$. All methods face the same simulated quadrotor, subject to a persistent wind disturbance: a steady force of $[0.35,-0.15,0]^\top$\,N superimposed with sinusoidal gusts of amplitude $[0.20,0.25,0]^\top$\,N, frequencies $[0.05,0.08,0]^\top$\,Hz, and phases $[0,\pi/3,0]^\top$.

Five methods are compared. TPC is the nominal tracking MPC with a prediction horizon $N=50$. CW-Reg is a control-weighted regularization baseline. DiffCtrl (Differentiable Control \cite{amos2018}) backpropagates the tracking cost through the MPC optimization layer to jointly update the model and controller parameters. F-SPC (Fixed-Surrogate SPC) adopts the same horizon as TPC but injects a preview of the estimated wind disturbance into the position-level prediction model and the feedforward tilt reference. U-SPC (Updated-Surrogate SPC) extends F-SPC by recomputing the auxiliary evaluation parameter $\theta_{\mathrm{emp}}$ every $\tau=10$ iterations to tighten surrogate--deployment alignment. Tracking RMSE is the Euclidean distance between tracked and reference positions along the closed-loop rollout. We additionally report aggregate control effort and one-step prediction MSE. All results are averaged over 3 random seeds, with standard deviations reported.

\subsection{Results}

\begin{figure}[t]
\centering
\includegraphics[width=0.5\linewidth]{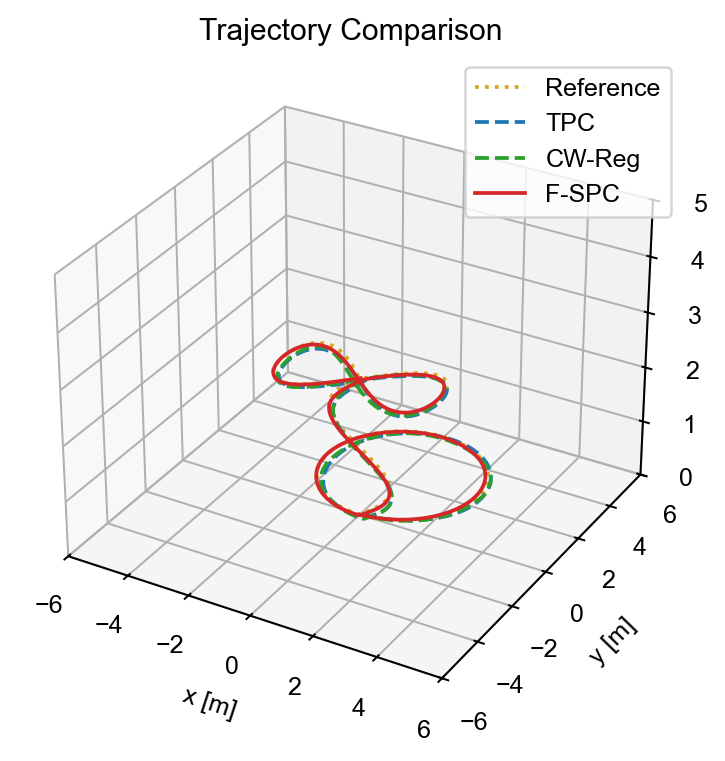}
\caption{Wind-disturbed quadrotor trajectory tracking.}
\label{fig:quadrotor_comparison}
\end{figure}

Figure~\ref{fig:quadrotor_comparison} displays the tracked trajectories. Both TPC and CW-Reg deviate visibly from the reference during the wind-disturbed maneuver, with the largest discrepancies appearing around the descent phase and the transition into the figure-8. The two SPC variants remain consistently closest to the reference throughout the rollout.

Table~\ref{tab:quadrotor_main} summarizes all metrics (mean $\pm$ std over 3 seeds). U-SPC achieves the best tracking RMSE ($0.0597 \pm 0.0039$\,m) and closed-loop cost ($16.5 \pm 0.6$), achieving reductions of $70\%$ and $42\%$ relative to TPC. F-SPC and DiffCtrl attain comparable tracking accuracy, both substantially outperforming TPC. A particularly instructive pattern emerges from the remaining columns: TPC produces the lowest prediction MSE but fails to deliver the best tracking. This contrast underscores the misalignment between prediction accuracy and downstream control quality—minimizing prediction error alone does not ensure good closed-loop behavior, and naively grafting control-relevant penalties onto the prediction loss can even worsen performance. The SPC variants accept modestly higher prediction error in exchange for markedly better closed-loop outcomes, which is precisely the motivation for a control-aware surrogate.

\begin{table}[t]
\caption{Results for the wind-disturbed quadrotor tracking experiment (mean $\pm$ std over 3 seeds).}
\label{tab:quadrotor_main}
\centering
\setlength{\tabcolsep}{1.6pt}
\renewcommand{\arraystretch}{1.05}
\footnotesize
\begin{tabular}{lccccc}
\toprule
\textbf{Metric} & \textbf{TPC} & \textbf{CW-Reg} & \textbf{DiffCtrl} & \textbf{F-SPC} & \textbf{U-SPC} \\
\midrule
RMSE\,(m) & .199\,{\tiny$\pm$\,.008} & .376\,{\tiny$\pm$\,.015} & .084\,{\tiny$\pm$\,.007} & .073\,{\tiny$\pm$\,.005} & \textbf{.060}\,{\tiny$\pm$\,.004} \\
CL Cost   & 28.7\,{\tiny$\pm$\,1.2} & 35.4\,{\tiny$\pm$\,1.8} & 19.2\,{\tiny$\pm$\,0.9} & 18.1\,{\tiny$\pm$\,0.7} & \textbf{16.5}\,{\tiny$\pm$\,0.6} \\
Ctrl eff. & 44.8\,{\tiny$\pm$\,0.9} & 37.2\,{\tiny$\pm$\,1.1} & 42.1\,{\tiny$\pm$\,0.8} & 43.3\,{\tiny$\pm$\,0.7} & 41.6\,{\tiny$\pm$\,0.6} \\
Pred MSE  & \textbf{.0015}\,{\tiny$\pm$\,.0002} & .0021\,{\tiny$\pm$\,.0003} & .0028\,{\tiny$\pm$\,.0003} & .0024\,{\tiny$\pm$\,.0002} & .0026\,{\tiny$\pm$\,.0003} \\
\bottomrule
\end{tabular}
\end{table}
\section{CONCLUSION}

We have proposed SPC, a control-aware surrogate refinement for system identification. For the fixed-surrogate variant under unconstrained control, we proved projected-gradient convergence and derived a conditional transfer diagnostic linking surrogate improvement to deployment performance. On a wind-disturbed quadrotor task, U-SPC reduces tracking RMSE by $70\%$ and closed-loop cost by $42\%$ relative to TPC, while TPC's prediction-optimal model fails to translate into the best closed-loop behavior.

\textbf{Limitations and Future Work.} The updated-surrogate variant remains a heuristic without formal guaranties. The bias constant $L_B$ is not computable from data alone, motivating the development of data-driven bounds. The disturbance scenarios are derived from a single initial model and held fixed throughout refinement, which may limit adaptability when the initial fit is poor; exploring iterative disturbance reestimation is a promising direction. 

\section*{Acknowledgment}
Claude was used to assist with the language editing of this manuscript.

\appendix
\section{Proofs for Section~\ref{sec:theory}}
\label{app:theory_proofs}

This appendix collects the proofs for the regularity, convergence, and conditional-transfer results stated in Section~\ref{sec:theory}.

\subsection{A Rollout Regularity Lemma}

\begin{lemma}[Rollout smoothness]
\label{lem:rollout_smoothness}
Fix a horizon $T\in\mathbb{N}$ and an open set $\mathcal O\subseteq\mathbb{R}^p$. Suppose
$f:\mathbb{R}^n\times\mathbb{R}^m\times\mathcal O\to\mathbb{R}^n$
is $C^3$. For any fixed initial state $x_0\in\mathbb{R}^n$ and disturbance sequence $W=(w_0,\dots,w_{T-1})$, define the trajectory recursively by
$x_{t+1}(U,\theta)=f(x_t(U,\theta),u_t;\theta)+w_t$, $x_0(U,\theta)=x_0$.
Then each map $(U,\theta)\mapsto x_t(U,\theta)$ is $C^3$, and consequently, each $F_i(U,\theta)$ is $C^3$.
\end{lemma}

\begin{proof}
We proceed by induction on $t$. The base case $x_0(U,\theta)=x_0$ is constant; hence, $C^3$. For the inductive step, suppose $(U,\theta)\mapsto x_t(U,\theta)$ is $C^3$. Then
\begin{equation}
\label{eq:app_rollout_induction}
x_{t+1}(U,\theta)=f\bigl(x_t(U,\theta),\,u_t;\,\theta\bigr)+w_t,
\end{equation}
which is a composition of $C^3$ maps (the induction hypothesis for $x_t$, the assumption on $f$, and a constant translation by $w_t$); hence, $C^3$. Since $J$ in \eqref{eq:cost_theory} is a sum of quadratic forms in $x_t$ and $u_t$, and $F_i$ is a finite average of such terms, both inherit $C^3$ regularity.
\end{proof}

\subsection{Proof of Proposition~\ref{prop:smoothness_revised}}

\begin{proof}
Fix an index $i$.

\paragraph*{Step 1: Existence and uniqueness of the optimizer.} Let $\theta\in\Theta$ be arbitrary. By Assumption~\ref{as:Fi_reg_revised}, the map $U\mapsto F_i(U,\theta)$ is continuous and $\mu_U$-strongly convex on $\mathbb{R}^{mT}$.

Strong convexity gives, for every $U\in\mathbb{R}^{mT}$,
\begin{equation}
\label{eq:app_strong_convexity}
F_i(U,\theta)
\ge
F_i(0,\theta)
+
\nabla_UF_i(0,\theta)^\top U
+
\frac{\mu_U}{2}\|U\|^2.
\end{equation}
Because the right-hand side grows without bound as $\|U\|\to\infty$, the function $F_i(\cdot,\theta)$ is coercive. A continuous coercive function on $\mathbb{R}^{mT}$ attains its minimum, so at least one minimizer exists. Strong convexity ensures uniqueness. Denote this minimizer by $U_i^*(\theta)$.

\paragraph*{Step 2: $C^2$ regularity of the optimizer map.} Define
\[
\Psi_i(U,\theta)\triangleq \nabla_U F_i(U,\theta).
\]
Since $F_i$ is $C^3$ on $\mathbb{R}^{mT}\times\mathcal O$, the map $\Psi_i$ is $C^2$ on that set. By the first-order optimality condition,
\[
\Psi_i(U_i^*(\theta),\theta)=0.
\]
Moreover,
\[
D_U\Psi_i(U_i^*(\theta),\theta) = \nabla_{UU}^2F_i(U_i^*(\theta),\theta).
\]
Assumption~\ref{as:Fi_reg_revised} yields
\[
\nabla_{UU}^2F_i(U_i^*(\theta),\theta)\succeq \mu_UI,
\]
so this matrix is invertible for every $\theta\in\Theta$.

The implicit function theorem then provides a neighborhood $\mathcal N_\theta\subseteq\mathcal O$ of $\theta$ and a unique $C^2$ map $\phi_\theta:\mathcal N_\theta\to\mathbb{R}^{mT}$ satisfying
\[
\Psi_i(\phi_\theta(\vartheta),\vartheta)=0 \qquad \forall \vartheta\in\mathcal N_\theta.
\]
Uniqueness of the optimizer forces $\phi_\theta(\vartheta)=U_i^*(\vartheta)$ on $\mathcal N_\theta$. Because these local representations agree on overlaps, $\theta\mapsto U_i^*(\theta)$ is globally $C^2$ on $\Theta$.

Differentiating the first-order condition
\[
\nabla_UF_i(U_i^*(\theta),\theta)=0
\]
with respect to $\theta$ yields
\[
\nabla_{UU}^2F_i(U_i^*(\theta),\theta)\,DU_i^*(\theta) + \nabla_{U\theta}^2F_i(U_i^*(\theta),\theta) = 0.
\]
Invertibility of $\nabla_{UU}^2F_i(U_i^*(\theta),\theta)$ gives
\[
DU_i^*(\theta) = - \Bigl[\nabla_{UU}^2F_i(U_i^*(\theta),\theta)\Bigr]^{-1} \nabla_{U\theta}^2F_i(U_i^*(\theta),\theta).
\]

\paragraph*{Step 3: $C^2$ regularity of the surrogate.} For fixed $\theta_{\mathrm{emp}}\in\Theta$, consider the two maps
\[
\theta\mapsto F_i(U_i^*(\theta),\theta), \qquad \theta\mapsto F_i(U_i^*(\theta),\theta_{\mathrm{emp}}).
\]
Since $U_i^*$ is $C^2$ and $F_i$ is $C^3$, both are $C^2$ by the chain rule. The surrogate $\widetilde{\mathcal L}(\cdot;\theta_{\mathrm{emp}})$, being a finite average of linear combinations of such maps, is itself $C^2$ on $\Theta$.

\paragraph*{Step 4: Explicit gradient formula.} The chain rule gives
\begin{align}
\nabla_\theta F_i(U_i^*(\theta),\theta)
&= \partial_\theta F_i(U_i^*(\theta),\theta) \notag\\
&\quad + \partial_UF_i(U_i^*(\theta),\theta)\,DU_i^*(\theta).
\end{align}
Because $U_i^*(\theta)$ minimizes $F_i(\cdot,\theta)$, the first-order condition $\partial_UF_i(U_i^*(\theta),\theta)=0$ yields the envelope-type identity
\begin{equation}
\label{eq:app_envelope}
\nabla_\theta F_i(U_i^*(\theta),\theta)
=
\partial_\theta F_i(U_i^*(\theta),\theta).
\end{equation}
For the evaluation term, an analogous chain-rule expansion gives
\begin{equation}
\label{eq:app_eval_chain}
\begin{aligned}
\nabla_\theta F_i(U_i^*(\theta),\theta_{\mathrm{emp}})
&= \partial_\theta F_i(U_i^*(\theta),\theta_{\mathrm{emp}}) \\
&\quad + \partial_U F_i(U_i^*(\theta),\theta_{\mathrm{emp}})
\, D U_i^*(\theta).
\end{aligned}
\end{equation}
Here the situation differs from the self-evaluation case: $U_i^*(\theta)$ minimizes $F_i(\cdot,\theta)$, not $F_i(\cdot,\theta_{\mathrm{emp}})$, so $\partial_UF_i(U_i^*(\theta),\theta_{\mathrm{emp}})\neq 0$ in general. However, the first term vanishes because the differentiation is with respect to $\theta$ while $\theta_{\mathrm{emp}}$ is held fixed, giving $\partial_\theta F_i(U_i^*(\theta),\theta_{\mathrm{emp}})=0$. Thus
\begin{equation}
\label{eq:app_fixed_eval}
\nabla_\theta F_i(U_i^*(\theta),\theta_{\mathrm{emp}})
=
\partial_UF_i(U_i^*(\theta),\theta_{\mathrm{emp}})\,DU_i^*(\theta).
\end{equation}

\paragraph*{Step 5: Lipschitz continuity of the gradient.} Continuity of $U_i^*$ together with compactness of $\Theta$ ensures that the image
\[
\mathcal M_i\triangleq \{U_i^*(\theta):\theta\in\Theta\} \subset \mathbb{R}^{mT}
\]
is compact, and hence so is
\[
\mathcal K_i \triangleq \{(U_i^*(\theta),\theta):\theta\in\Theta\} \subset \mathbb{R}^{mT}\times\mathcal O.
\]

All partial derivatives of $F_i$ up to order three are continuous and therefore bounded on $\mathcal K_i$. Furthermore, the uniform positive-definiteness of $\nabla_{UU}^2F_i$ yields
\[
\left\| \bigl[\nabla_{UU}^2F_i(U_i^*(\theta),\theta)\bigr]^{-1} \right\| \le \frac{1}{\mu_U}, \qquad \forall \theta\in\Theta.
\]\label{eq:app_surrogate_grad}
By the formula for $DU_i^*(\theta)$, the Jacobian $DU_i^*$ is continuous and bounded on $\Theta$. Because $U_i^*$ is $C^2$, every term arising from differentiating \eqref{eq:app_surrogate_grad} once more is continuous on $\Theta$. Compactness then guaranties a finite constant $L_\nabla$ satisfying
\[
\sup_{\theta\in\Theta} \|\nabla^2\widetilde{\mathcal L}(\theta;\theta_{\mathrm{emp}})\| \le L_\nabla.
\]
An application of the mean value theorem completes the argument:
\[
\begin{aligned}
&\|\nabla \widetilde{\mathcal L}(\theta;\theta_{\mathrm{emp}})
-\nabla \widetilde{\mathcal L}(\theta';\theta_{\mathrm{emp}})\|\\
&\qquad \le L_\nabla\|\theta-\theta'\|,
\qquad \forall \, \theta,\theta'\in\Theta.
\end{aligned}
\]
\end{proof}

\subsection{A Projection Lemma}

\begin{lemma}[Projection optimality]
\label{lem:projection_optimality}
Let $\Theta\subseteq\mathbb{R}^p$ be nonempty, closed, and convex. For any $y\in\mathbb{R}^p$, let $\Pi_\Theta(y)$ denote the Euclidean projection of $y$ onto $\Theta$. Then $\bar\theta=\Pi_\Theta(y)$ if and only if
\[
\langle \bar\theta-y, z-\bar\theta\rangle \ge 0, \qquad \forall z\in\Theta.
\]
Equivalently,
\[
\langle y-\bar\theta, z-\bar\theta\rangle \le 0, \qquad \forall z\in\Theta.
\]
\end{lemma}

\begin{proof}
This is a standard result in convex analysis; see, e.g., Nesterov \cite{nesterov2018}.
\end{proof}

\subsection{Proof of Theorem~\ref{thm:stationarity_revised}}

\begin{proof}
Write $h(\theta)\triangleq \widetilde{\mathcal L}(\theta;\theta_{\mathrm{emp}})$. Since $\nabla h$ is $L_\nabla$-Lipschitz by Proposition~\ref{prop:smoothness_revised}, the standard descent lemma gives
\begin{equation}
\label{eq:app_sufficient_decrease}
\begin{aligned}
h(\theta^{(k+1)})
\le\;& h(\theta^{(k)})
+ \langle \nabla h(\theta^{(k)}),\,\theta^{(k+1)}-\theta^{(k)}\rangle \\
&+ \frac{L_\nabla}{2}\|\theta^{(k+1)}-\theta^{(k)}\|^2.
\end{aligned}
\end{equation}
Substituting the projected-gradient update $\theta^{(k+1)}=\Pi_\Theta[\theta^{(k)}-\eta\nabla h(\theta^{(k)})]$ and applying the projection optimality condition (Lemma~\ref{lem:projection_optimality}) with the step size $\eta\le 1/L_\nabla$ yields the per-iteration descent
\begin{equation}
\label{eq:app_descent_gm}
h(\theta^{(k+1)})
\le
h(\theta^{(k)})-\frac{\eta}{2}\|\mathcal G_\eta(\theta^{(k)})\|^2.
\end{equation}
Summing from $k=0$ to $K-1$ and using $h(\theta^{(K)})\ge h^*$ produces
\begin{equation}
\label{eq:app_telescope}
\frac{\eta}{2}\sum_{k=0}^{K-1}\|\mathcal G_\eta(\theta^{(k)})\|^2
\le h(\theta^{(0)})-h(\theta^{(K)})
\le h(\theta^{(0)})-h^*.
\end{equation}
Since the minimum of a finite set is at most its average,
\[
\min_{0\le k\le K-1}\|\mathcal G_\eta(\theta^{(k)})\|^2
\le \frac{2(h(\theta^{(0)})-h^*)}{\eta K}.
\]
Furthermore, \eqref{eq:app_telescope} implies $\sum_{k=0}^{\infty}\|\mathcal G_\eta(\theta^{(k)})\|^2<\infty$, so $\|\mathcal G_\eta(\theta^{(k)})\|\to 0$. Since $\Theta$ is compact and $\mathcal G_\eta$ continuous, every accumulation point $\bar\theta$ satisfies $\mathcal G_\eta(\bar\theta)=0$, i.e., $\langle \nabla h(\bar\theta),\,z-\bar\theta\rangle \ge 0$ for all $z\in\Theta$.
\end{proof}

\subsection{Proof of Proposition~\ref{prop:bias_revised}}
\begin{proof}
By definition,
\(
V(\theta)=\widetilde{\mathcal L}(\theta;\theta_{\mathrm{emp}})
+ B(\theta;\theta_{\mathrm{emp}})
\).
Evaluating the same identity at \(\theta'\) and subtracting yields
\[
\begin{aligned}
V(\theta)-V(\theta')
={}&\widetilde{\mathcal L}(\theta;\theta_{\mathrm{emp}})
-\widetilde{\mathcal L}(\theta';\theta_{\mathrm{emp}}) \\
&+ B(\theta;\theta_{\mathrm{emp}})
- B(\theta';\theta_{\mathrm{emp}}).
\end{aligned}
\]
\end{proof}

\subsection{Proof of Theorem~\ref{thm:transfer_revised}}

\begin{proof}
Proposition~\ref{prop:bias_revised} gives
\[
\begin{aligned}
V(\theta^{(K)})-V(\theta^{(0)})
={}&\widetilde{\mathcal L}(\theta^{(K)};\theta_{\mathrm{emp}})
-\widetilde{\mathcal L}(\theta^{(0)};\theta_{\mathrm{emp}})\\
&+B(\theta^{(K)};\theta_{\mathrm{emp}})-B(\theta^{(0)};\theta_{\mathrm{emp}}).
\end{aligned}
\]
Assumption~\ref{as:bias_reg_revised} bounds the bias variation:
\[
B(\theta^{(K)};\theta_{\mathrm{emp}}) - B(\theta^{(0)};\theta_{\mathrm{emp}}) \le L_B\|\theta^{(K)}-\theta^{(0)}\|.
\]
Hence
\[
\begin{aligned}
V(\theta^{(K)})-V(\theta^{(0)})
\le{}&-\Bigl(\widetilde{\mathcal L}(\theta^{(0)};\theta_{\mathrm{emp}})
-\widetilde{\mathcal L}(\theta^{(K)};\theta_{\mathrm{emp}})\Bigr)\\
&+L_B\|\theta^{(K)}-\theta^{(0)}\|.
\end{aligned}
\]
When the surrogate decrease exceeds the bias perturbation,
\[
\widetilde{\mathcal L}(\theta^{(0)};\theta_{\mathrm{emp}}) - \widetilde{\mathcal L}(\theta^{(K)};\theta_{\mathrm{emp}}) > L_B\|\theta^{(K)}-\theta^{(0)}\|,
\]
the right-hand side is strictly negative and therefore $V(\theta^{(K)})<V(\theta^{(0)})$.
\end{proof}

\section*{Acknowledgment}
Claude was used to assist with the language editing of this manuscript.
\bibliographystyle{IEEEtran}
\bibliography{references}

@article{bristow2006,
  title={A survey of iterative learning control: A learning-based method for highperformance tracking control},
  author={Tharayil, DABM and Alleyne, Andrew G},
  journal={IEEE Control systems magazine},
  volume={26},
  number={3},
  pages={96--114},
  year={2006}
}

@article{gevers2005,
  title={Identification for control: From the early achievements to the revival of experiment design},
  author={Gevers, Michel},
  journal={European journal of control},
  volume={11},
  number={4-5},
  pages={335--352},
  year={2005},
  publisher={Elsevier}
}

@article{hjalmarsson2005,
  title={From experiment design to closed-loop control},
  author={Hjalmarsson, H{\aa}kan},
  journal={Automatica},
  volume={41},
  number={3},
  pages={393--438},
  year={2005},
  publisher={Elsevier}
}

@article{formentin2021,
  title={Control-oriented regularization for linear system identification},
  author={Formentin, Simone and Chiuso, Alessandro},
  journal={Automatica},
  volume={127},
  pages={109539},
  year={2021},
  publisher={Elsevier}
}

@inproceedings{bansal2017,
  title={Goal-driven dynamics learning via Bayesian optimization},
  author={Bansal, Somil and Calandra, Roberto and Xiao, Ted and Levine, Sergey and Tomlin, Claire J},
  booktitle={2017 IEEE 56th Annual Conference on Decision and Control (CDC)},
  pages={5168--5173},
  year={2017},
  organization={IEEE}
}

@article{amos2018,
  title={Differentiable mpc for end-to-end planning and control},
  author={Amos, Brandon and Jimenez, Ivan and Sacks, Jacob and Boots, Byron and Kolter, J Zico},
  journal={Advances in neural information processing systems},
  volume={31},
  year={2018}
}

@article{drgona2022,
  title={Differentiable predictive control: Deep learning alternative to explicit model predictive control for unknown nonlinear systems},
  author={Drgo{\v{n}}a, J{\'a}n and Ki{\v{s}}, Karol and Tuor, Aaron and Vrabie, Draguna and Klau{\v{c}}o, Martin},
  journal={Journal of Process Control},
  volume={116},
  pages={80--92},
  year={2022},
  publisher={Elsevier}
}

@inproceedings{romero2024,
  title={Actor-critic model predictive control},
  author={Romero, Angel and Song, Yunlong and Scaramuzza, Davide},
  booktitle={2024 IEEE International Conference on Robotics and Automation (ICRA)},
  pages={14777--14784},
  year={2024},
  organization={IEEE}
}

@article{brunke2022,
  title={Safe learning in robotics: From learning-based control to safe reinforcement learning},
  author={Brunke, Lukas and Greeff, Melissa and Hall, Adam W and Yuan, Zhaocong and Zhou, Siqi and Panerati, Jacopo and Schoellig, Angela P},
  journal={Annual Review of Control, Robotics, and Autonomous Systems},
  volume={5},
  number={1},
  pages={411--444},
  year={2022},
  publisher={Annual Reviews}
}

@article{elmachtoub2022,
  title={Smart “predict, then optimize”},
  author={Elmachtoub, Adam N and Grigas, Paul},
  journal={Management Science},
  volume={68},
  number={1},
  pages={9--26},
  year={2022},
  publisher={INFORMS}
}

@article{cui2025,
  title={A “Smart Model-then-Control” Strategy for the Scheduling of Thermostatically Controlled Load},
  author={Cui, Xueyuan and Liu, Boyuan and Li, Yehui and Wang, Yi},
  journal={IEEE Transactions on Smart Grid},
  year={2025},
  publisher={IEEE}
}

@inproceedings{favaro2025,
  title={Decision-Focused Learning for Complex System Identification: HVAC Management System Application},
  author={Favaro, Pietro and Toubeau, Jean-Fran{\c{c}}ois and Vall{\'e}e, Fran{\c{c}}ois and Dvorkin, Yury},
  booktitle={Proceedings of the 16th ACM International Conference on Future and Sustainable Energy Systems},
  pages={347--358},
  year={2025}
}

@article{lin2025,
  title={Maximizing the Value of Predictions in Control: Accuracy Is Not Enough},
  author={Lin, Yiheng and Yeh, Christopher and Chen, Zaiwei and Wierman, Adam},
  journal={arXiv preprint arXiv:2506.04497},
  year={2025}
}

@article{yu2020,
  title={The power of predictions in online control},
  author={Yu, Chenkai and Shi, Guanya and Chung, Soon-Jo and Yue, Yisong and Wierman, Adam},
  journal={Advances in Neural Information Processing Systems},
  volume={33},
  pages={1994--2004},
  year={2020}
}

@book{nesterov2018,
  title={Lectures on convex optimization},
  author={Nesterov, Yurii and others},
  volume={137},
  year={2018},
  publisher={Springer}
}

\end{document}